\documentclass[prd,twocolumn,amsmath,amssymb,floatfix,superscriptaddress]{revtex4-1}

\usepackage{graphicx}
\usepackage{amssymb}
\usepackage{amsmath}
\usepackage{color}
\usepackage{ wasysym }
\usepackage{hyperref}
 \usepackage{tabu}

\def\barray{\begin{array}}
\def\earray{\end{array}}
\def\be{\begin{equation}}
\def\ee{\end{equation}}
\def\ben{\begin{equation} \nonumber}
\def\een{\end{equation}}
\def\ban{\begin{eqnarray*}}
\def\ean{\end{eqnarray*}}
\def\ba{\begin{eqnarray}}
\def\ea{\end{eqnarray}}

\def\({\left(}
\def\){\right)}

\def\apj{ApJ}

\def\mnras{MNRAS}

\def\aj{AJ}

\def\physrep{Phys. Rep.}

\graphicspath{{./fig/}}

\begin{document}

\title{Can Holographic dark energy models fit the observational data?}

\author{Mohammad Malekjani}
\affiliation{Department of Physics, Bu-Ali Sina University, Hamedan
65178, 016016, Iran}

\author{Mehdi Rezaei}
\email{rezaei@irimo.ir}
\affiliation{Research Institute for Astronomy and Astrophysics of Maragha (RIAAM)
, Maragha, Iran, P.O.Box:55134-441}
\affiliation{Iran Meteorological Organization, Hamedan Research Center for Applied
Meteorology, Hamedan, Iran}

\author{Iman A. Akhlaghi}
\affiliation{Department of Physics, Institute for advanced studies in Basic Sciences, Zanjan 45137-66731, Iran}

\begin{abstract}
    In this work we investigate the holographic dark energy models with slowly time-varying model parameter defined based on the current Hubble horizon length scale. While the previous studies on the three popular holographic dark energy models defined based on the future event horizon, Ricci scale and Granda-Oliveros IR cutoffs showed that these models cannot fit the observational data \citep{Akhlaghi:2018knk}, in this work we show that the holographic dark energy models with time-varying model parameter  defined on the current Hubble radius are well favored by observations. Using the standard $\chi^2$ minimization in the context of Markov Chain Monte Carlo method, we compare the ability of holographic dark energy models with time-varying $c^2$ parameter constructed on the current Hubble length scale against different sets of observational data namely expansion data, growth rate data and expansion+growth rate data respectively. Based on the values of  Akaike and Bayesian information criteria, we find that these types of holographic dark energy models are well fitted to both expansion and growth rate observations as equal to $\Lambda$CDM cosmology. We also put constraints on the cosmological parameters and show that the transition epoch form early decelerated to current accelerated expansion calculated in holographic dark energy models with time-varying model parameter defined on the Hubble length is consistent with observations.
 
\end{abstract}
\maketitle

\section{Introduction}
The current accelerated expansion of the universe can be well explained either by introducing an exotic cosmic fluid with a sufficiently negative pressure dubbed dark energy (DE), or by modifying the
standard theory of gravity on extragalactic scales \citep{Riess1998,Perlmutter1999,Kowalski2008}. A combined analysis of cosmological observations indicates that the current universe is spatially flat and dark energy occupies about $2/3$ of the total energy budget of it \citep{Bennett:2003bz,Spergel:2003cb,Peiris:2003ff}.
The first theoretical candidate of dark energy is the well known cosmological constant $\Lambda$ in which the equation of state (EoS) parameter is equal to $-1$.  Although the concordance $\Lambda$ cosmology is consistent with the cosmological observations, it always suffers from two puzzles: the fine-tuning and the cosmic coincidence \citep{Weinberg1989,Padmanabhan2003,Copeland2006}. Alternatively, in the last two decades, numerous other candidates for dark energy with a time evolving energy density have also been proposed in the literatures in order to solve or at least alleviate the above cosmological problems  \citep{Caldwell:1997ii,Armendariz2001,Caldwell2002,Elizalde:2004mq}. In these models, the EoS parameter of dark energy varies a function of cosmic redshift. Unfortunately, most of the proposed models for DE are phenomenological and the nature of DE is unknown. Therefore, some cosmologists were motivated to propose a model in which the origin of DE is based on physical principles,
namely it is related with the effects of quantum gravity. In this regard, the first model was proposed by \cite{Li:2004rb} by applying the holographic principle, which is the fundamental principle in the quantum gravity scenario \citep{tHooft1993,Susskind1995}, on cosmological scale to propose a model for DE called the holographic dark energy model. The holographic principle indicates that all physical information inside
in a space-volume can be interpreted as a hologram which corresponds to a theory locating on the boundary of that space \cite{tHooft1993,Susskind1995}. It has been shown that the effective local quantum field theories
greatly over-count degrees of freedom because the entropy  $S$, in a box of size $L$ with UV cut-off $\Lambda_c$ scales extensively for an effective quantum field theory, $S\sim L^3\Lambda_c^3$ \cite{tHooft1993,Susskind1995}. Historically, the peculiar thermodynamics of black hole has led Bekenstein to postulate that the maximum entropy in a box of volume $L^3$ behaves non extensively, growing only as the area of the box,
i.e. there is a so-called Bekenstein entropy bound, $S \leq S_{BH} \equiv\pi M_p^2L^2$ \citep{Bekenstein:1973ur,Bekenstein:1974ax}. This non-extensive scaling indicates that quantum field theory breaks down in large volume. To conciliate this
breakdown with the success of local quantum field theory in describing observed particle
phenomenology, Cohen et al.\citep{Cohen1999} suggested a more restrictive bound, the energy bound.
They pointed out that in quantum field theory a short distance (UV) cutoff is related to a
long distance (IR) cutoff due to the limit set by forming a black hole. In other words, If we have a system with size $L$, its total energy should not exceeds the mass of a black hole with the same size, i.e., $L^3\rho _\Lambda\leq L M_{\rm p}^2$, where $\rho _\Lambda$ is the quantum zero-point
energy density caused by UV cutoff.
In cosmological contexts, when the whole of the universe is taken into account, the vacuum
energy related to the holographic principle can be viewed as holographic dark energy (HDE) with
energy density given by

\begin{eqnarray}\label{hde}
\rho_{\rm d}=3 c^2M_{\rm p}^2L^{-2}\;.
\end{eqnarray}

where $c^2$ is a dimensionless numerical parameter and the coefficient $3$ is for convenience. Very often, for the sake of simplicity, the $c^2$ parameter is assumed constant. Depending on length scale $L$, we can define the following types of HDE models \citep[see also][]{Bousso:1999xy,Horava2000,Thomas2002,Shen:2004ck,Enqvist:2004ny,Huang:2004ai,Gao:2007ep,Duran:2010ky,Radicella:2010vf,Sheykhi:2011cn,Mehrabi:2015kta}. ({\it i}) The simplest choice is the Hubble length, i.e., $L = H^{-1}$. In this case the density of DE will be close to the observational data, but the current accelerated expansion of the universe cannot be recovered \citep{Horava2000,Thomas2002,Pavon2005}. ({\it ii}) The other choice is particle horizon. In this case, it is impossible for HDE model to provide an accelerated expansion of the universe \citep{Li:2004rb}. ({\it iii}) Another simple choice for $L$ is the future event horizon \citep{Li:2004rb, Wang:2005ph}. It has been shown that this HDE model accommodates the
late time accelerated expansion \citep{Pavon2005,Zimdahl2007}. Also, the coincidence and the fine-tuning problems are well
alleviated at this length scale \citep{Li:2004rb}. The HDE model defined on the basis of future event horizon has been extensively investigated in recent years \citep{Huang:2004wt,Kao:2005xp,Zhang:2005hs,Wang:2005ph,Chang:2005ph,Zhang:2007sh,Micheletti:2009jy,Xu:2012aw,Zhang:2013mca,Li:2013dha,Zhang:2014ija,Zhang:2015rha}. ({\it iv}) The length scale $L$ can be chosen as the curvature of spacetime, namely the Ricci scalar $R$ 
\citep{Gao:2007ep,Zhang:2009un}.  It has been shown that the Ricci HDE model is consistent
with the observations of supernova type Ia \citep{Zhang:2009un,Easson:2010av}. ({\it v}) Another choice for length scale $L$ is proposed by Granda \& Oliveros (GO) \citep{Granda:2008dk}. GO cut-off defined based on the combination of the Hubble parameter together with its time derivative. In this proposal the late time accelerated phase of expansion is well achieved \citep{Granda:2008dk} and also this model is consistent with supernova type Ia observations \citep{Wang:2016och}. Except cases ({\it i \& ii}) in which the current accelerated phase of expansion is not achieved, other cases ,i.e., cases ({\it iii, iv \& v}), are the most popular types of HDE models studied extensively in the literature. Recently, it has been shown that the HDE models defined based on future even horizon, Ricci scale and GO cut-off, i.e., cases ({\it iii, iv \& v}), are fitted to observational growth rate data in perturbation level as equally as concordance $\Lambda$CDM universe \citep{Akhlaghi:2018knk}. In addition, as was mentioned above, these models are consistent with the observations of SNIa. However, by combination of all observational data in expansion level including  SnIa, baryonic acoustic oscillation (BAO), cosmic microwave background (CMB) radiation, Big Bang Nucluesenthsis (BBN) and Hubble expansion data, \cite{Akhlaghi:2018knk} showed that the Ricci and GO HDE models have a strong tension with observations and consequently disfavored \citep[see also][]{Wang:2016och}. This result is unchanged when we combine all observational data in expansion and perturbation levels \citep{Akhlaghi:2018knk}. Moreover, the HDE model defined based on event horizon, i.e., case ({\it iii}), is in a mild tension with expansion and expansion+growth rate data \citep{Akhlaghi:2018knk}. Also from theoretical point of view, since the HDE model ({\it iii}) is defined based on the event horizon length  scale, an obvious  drawback concerning causality appears in this scenario. The above statements motivate us to consider another possibility for the definition of HDE model. In a more general case, we can consider $c^2$ parameter in Eq.(\ref{hde}) as a slowly varying function of time. In fact there are no strong evidences supporting that $c^2$ parameter should be constant. Adopting the time slowly varying of $c^2$, we can reconsider the first choice, i.e. $H^{-1}$, as an IR cut-off for length scale $L$. Interestingly, it has been shown that the HDE model with slowly varying $c^2$ term defined in Hubble length $H^{-1}$ can simultaneously drive accelerated expansion and solve the coincidence problem  \citep[see][]{Pavon2005,Pavon:2006qm,Guberina:2006qh,Radicella:2010vf}.  Also the significant advantage of this model is that the density of HDE  is not depending on the future or the past evolution of the universe.

In this paper we investigate the HDE models defined in Hubble length with varying $c^2$ term in both expansion and perturbation levels. We first examine the model against the latest observational data in expansion level including those of SNIa, BAO, CMB shift parameter, BBN and expansion Hubble data. In next step, we study the model in perturbation level using the latest growth rate data, i.e., $f(z)\sigma_8$ data. In fact, DE not only accelerates the expansion rate of the universe but also changes the evolution of matter perturbations and consequently the formation epochs of large scale structures of the universe \citep{ArmendarizPicon:2000dh,Tegmark:2003ud,Pace2010}. 
Moreover, in the case of dynamical DE models, in which the EoS parameter evolves with cosmic time, the growth of large scale structures are also affected by perturbations of DE \citep{Hu:2004yd,Ballesteros:2008qk,mota3,Basilakos:2010fb,Sapone:2012nh,Basse:2013zua,Pace:2013pea,Batista:2014uoa,Basilakos:2014yda,Pace:2014taa,Nesseris:2014mfa,Malekjani:2015pza,Mehrabi:2015kta,Malekjani:2016edh,Rezaei:2017hon}. In the context of Markov Chain Monte Carlo (MCMC) analysis, we setup an extended formalism in which the background expansion data are joined with the growth rate data to put tight constraints on the parameters of HDE model and evaluate it against combined observational data \citep[for similar studies, see][]{Cooray:2003hd,Corasaniti:2005pq,Blake:2011rj,Nesseris:2011pc,Basilakos:2012uu,Yang:2013hra,mota6,Contreras:2013bol,Chuang:2013hya,Li:2014mua,Mehrabi:2015hva,Basilakos:2016xob,mota7,Rivera:2016zzr,Rezaei:2017yyj}.

The paper is organized as follows. In Sect. \ref{sect:hde}, firstly, we introduce the FRW cosmology in the context of HDE models with varying $c^2$ parameter defined in Hubble length and secondly investigate four different parameterizations of $c^2$ parameter. In Sect.\ref{sect:bg}, we implement the likelihood analysis in the context of MCMC method using the geometrical expansion data to place constraints on the parameters of HDE models and compare the validity of models against observations. Then the evolution of main cosmological quantities in background level according to the dynamics of HDE models is studied. 
 In Sect.\ref{growth}, the growth of perturbations in our models is studied. We then perform another likelihood analysis using the solely growth rate data to examine the HDE models in perturbation level. Eventually, 
 we perform an overall likelihood analysis using the background + growth rate data to test the HDE models against the combined cosmological data. In Sect.\ref{conlusion}, the paper is concluded.

\section{HDE with varying $c^2$ term}\label{sect:hde}

Here we investigate the background evolution of HDE with the slowly varying $c^2$ term by taking in to account the current value of Hubble horizon as the IR cutoff \citep[see also][]{Pavon2005,Pavon:2006qm,Guberina:2006qh,Radicella:2010vf}. We are motivated to consider why one should consider the time varying function for $c^2$ parameter in HDE models defined in Hubble length scale. In fact in DE cosmologies, DE dominates the universe at very late times ($\rho_{\rm total}=\rho_{\rm d}$) . So from the Freidmann equation, we have $\rho_{\rm d}(z\rightarrow -1)=3M_{\rm p}^2H^2(z\rightarrow -1)$. On the other hand, from Eq.\ref{hde}, we can say $\rho_{\rm d}(z\rightarrow -1)=3c^2(z\rightarrow -1)M_{\rm p}^2H_0^2$. Hence we obtain the value of $c^2$ parameter at very late times as $c^2=H^2/H_0^2=E^2(z\rightarrow -1)$. It is obvious, at earlier times, $c^2$ cannot be equal to its value at far future ($E^2(z\rightarrow -1)$), because it would not leave room for dark matter and radiation. Therefore, to describe the evolution of energy densities at both  matter dominated and dark energy dominated epochs, time varying $c^2$ is inevitable. In order to satisfy the slowly varying condition, we shall use four wellknown parameterizations to describe $c(z)$. These parameterizations are Chevalier-Polarski-Linder (CPL) parameterization (Model 1) \citep{Chevallier2001}, Jassal-Bagla-Padmanabhan (JBP) parameterization (Model2) \citep{Jassal:2004ej}, Wetterrich parameterization (Model 3) \citep{Wetterich:2004pv}, and Ma-Zhang parameterization (Model 4) \citep{Ma:2011nc}, described in terms of redshift as follows:

\begin{eqnarray}\label{par1}
{\rm Molde (1):}~~& c(z)=c_0+c_1\frac{z}{1+z}\;.
\end{eqnarray}

\begin{eqnarray}\label{par2}
{\rm Model (2):}~~& c(z)=c_0+c_1\frac{z}{(1+z)^2}\;.
\end{eqnarray}

\begin{eqnarray}\label{par3}
{\rm Model (3):}~~& c(z)=\frac{c_0}{1+c_1\ln (1+z)}\;.
\end{eqnarray}

\begin{eqnarray}\label{par4}
{\rm Model (4):}~~& c(z)=c_0+c_1\left(  \frac{\ln (2+z)}{1+z}-\ln 2\right) \;.
\end{eqnarray}

One can see that in all of the above equations, setting $c_1=0$ leads to the original HDE model with constant $c$ parameter. Notice that the original HDE models in Hubble length cannot explain the acceleration of the universe \citep{Horava2000,Thomas2002,Hsu2004}.
For isotropic and homogeneous spatially flat FRW cosmologies, driven by radiation, 
non-relativistic pressure-less matter and DE, the first Friedmann equation can be written as

\begin{eqnarray}\label{frid1}
H^2=\frac{1}{3M^2_{\rm p}}(\rho_{\rm r}+\rho_{\rm m}+\rho_{\rm d})\;,
\end{eqnarray}
where $H\equiv {\dot a}/a$ is the Hubble parameter, $\rho_{\rm r}$, $\rho_{\rm m}$ and $\rho_{\rm d}$ are 
the energy densities of radiation, pressureless matter and DE, respectively. Introducing the density parameters $\Omega_{\rm i}$, for radiation, non-relativistic pressureless matter and DE, we obtain

\begin{eqnarray}\label{omega}
\Omega_{\rm r}=\frac{\rho_{\rm r}}{3M^2_{\rm p}H^2}~~& \Omega_{\rm m}=\frac{\rho_{\rm m}}{3M^2_{\rm p}H^2}&~~ \Omega_{\rm d}=\frac{\rho_{\rm d}}{3M^2_{\rm p}H^2}\;.
\end{eqnarray}

By using these parameters, we can write Eq.\ref{frid1} as
 
\begin{eqnarray}\label{omegasum}
\Omega_{\rm r}+ \Omega_{\rm m}+\Omega_{\rm d}=1\;.
\end{eqnarray}
In the absence of interactions among the three fluids the conservation equations for corresponding 
energy densities are given by
 \begin{eqnarray}\label{continuity}
 && \dot{\rho_{\rm r}}+4H\rho_{\rm r}=0\;,\label{radiation}\\
&&\dot{\rho_{\rm m}}+3H\rho_{\rm m}=0\;,\label{matter}\\
&&\dot{\rho_{\rm d}}+3H(1+w_{\rm d})\rho_{\rm d}=0\;\label{de},
 \end{eqnarray}
where the over-dot denotes a derivative with respect to cosmic time $t$. From Eq.\ref{hde} and using current value of Hubble parameter as IR cutoff the energy density of HDE models can be written as

\begin{eqnarray}\label{GHDE}
\rho_{\rm d}=3 c^2M^2_{\rm p}H^2_0\;.
\end{eqnarray}

Now using Eq.\ref{omega} we can obtain energy density for the HDE as

\begin{eqnarray}\label{omeghde}
\Omega_{\rm d}=\frac{c^2(z)}{E^2(z)}\;.
\end{eqnarray}
where $E(z)=H(z)/H_0$. Replacing Eq.\ref{omeghde} into Eq.\ref{omegasum} and using the evolution functions of $\Omega_{\rm r}$ and $\Omega_{\rm m}$ we arrive 

\begin{eqnarray}\label{e20}
\frac{\Omega^0_{\rm m}(1+z)^3}{E^2(z)}+\frac{\Omega^0_{\rm r}(1+z)^4}{E^2(z)}+\frac{c^2(z)}{E^2(z)}=1\;,
\end{eqnarray}
where $\Omega^0_{\rm m}$ and $\Omega^0_{\rm r}$ are the present values of matter and radiation density parameters respectively. Notice that we utilize $\Omega_{\rm r0}=2.469\times 10^{-5}h^{-2}(1.6903)$ to obtain the current value of the energy density of radiation, while $h=H_{0}/100$\citep{Hinshaw:2012aka}. Therefor the Hubble parameter, $E(z)$ takes the form
\begin{eqnarray}\label{e2}
E(z)=\sqrt{\Omega^0_{\rm m}(1+z)^3+\Omega^0_{\rm r}(1+z)^4+c^2(z)}\;.
\end{eqnarray}

If we take the time derivative of Eq.\ref{GHDE}, we obtain

\begin{eqnarray}\label{GHDEdot}
\dot{\rho}_{\rm d}=2\rho_{\rm d}\frac{\dot{c}(z)}{c(z)}\;,
\end{eqnarray}
Inserting Eq.\ref{GHDEdot} into Eq.\ref{de}, we can obtain the EoS parameter of HDE models as 

\begin{eqnarray}\label{EOS}
w_{\rm d}=-1-\frac{2}{3}\frac{c'(z)}{c(z)}\;,
\end{eqnarray}

where prime means derivative with respect to $x=\ln{a}$. Clearly, for constant $c$ parameter we have $c'(z)=0$ which leads to $w_{\rm d}=-1$ representing concordance $\Lambda$CDM universe. We see from Eq.\ref{EOS} that the evolution of EoS of HDE models depends on the evolution of $c(z)$. When we have $c(z)c'(z) < 0$, HDE models evolves in quintessence regime, i.e. $w_{\rm d} >-1$ and in other hand when $c(z)c'(z) > 0$, HDE evolves in phantom regime, i.e. $w_{\rm d} <-1$.  In what follow, we obtain the EoS parameter of HDE models using four different parameterizations of $c(z)$ introduced in Eqs.(\ref{par1}-\ref{par4}).

\subsection{Model (1), The CPL parameterization}
Based on Eq.\ref{par1},one can see that at early times $(z\rightarrow\infty)$, $c \rightarrow c_0+c_1$ while at the present time we have $z\rightarrow 0$ and thus $c\rightarrow c_0$. This means that parameter $c$ changes slowly from $c_0+c_1$ at early times to $c_0$ at present time. Taking derivatives of Eq. \ref{par1} with respect to $x=\ln a$ we obtain
\begin{eqnarray}\label{par1driv}
c'(z)=-\frac{c_1}{1+z}\;.
\end{eqnarray}

Replacing Eqs.(\ref{par1driv} \& \ref{par1}) in Eq.\ref{EOS}, the EoS parameter of model (1) takes the form
 
\begin{eqnarray}\label{EOS1}
w_{\rm d}=-1+\frac{2}{3}\frac{c_1}{c_0(1+z)+c_1z}\;,
\end{eqnarray}

\subsection{Model (2), The JBP parameterization}
In this case the parameterization for $c(z)$ is given by Eq.(\ref{par2}). Taking derivative of Eq.\ref{par2}, we have

\begin{eqnarray}\label{par2prim}
c'(z)=-c_1\frac{1-z}{(1+z)^2}\;.
\end{eqnarray}

Substituting Eqs. \ref{par2} and \ref{par2prim} into Eq. \ref{EOS}, we obtain the EoS parameter for model (2) as

\begin{eqnarray}\label{EOS2}
w_{\rm d}=-1+\frac{2}{3}\frac{c_1(1-z)}{c_0(1+z)^2+c_1z}\;.
\end{eqnarray}

In this form, at $z=0$ we have $c=c_0$ and in other hand at early times, when $z\rightarrow\infty$ we have $c=c_0$. But among these two epochs and also in the future, we could have $c\neq c_0$.

\subsection{Model (3), The Wetterich parameterization}
The other parametrization we consider in this work, is Wetterich-type which is given by Eq. (\ref{par3}). In this form, at present time, $z=0$, we have $c=c_0$ while at the early times where $z\rightarrow\infty$ we have $c=0$. Thus based on Eq. \ref{hde}, at the early universe the HDE model did not have considerable role in
the evolution of the universe. Like previous parameterizations in order to obtain $c'(z)$ and $w_{\rm d}$ we should take derivatives of Eq. \ref{par3} with respect to $\ln a$ which leads to

\begin{eqnarray}\label{par3prim}
c'(z)=\frac{c_0c_1}{\left( 1+c_1\ln(1+z)\right)^2}\;.
\end{eqnarray}
Substituting Eqs. \ref{par3} and \ref{par3prim} into Eq. \ref{EOS}, one gets the EoS parameter for model (3) as
\begin{eqnarray}\label{EOS3}
w_{\rm d}=\frac{2}{3}\frac{c_1\left( 1+c_1\ln(1+z)\right) }{c^2_0-\left( 1+c_1\ln(1+z)\right)^2}\;.
\end{eqnarray}

\subsection{Model (4), The Ma-Zhang parameterization}

The last parameterization we choose for $c(z)$ is the Ma-Zhang parameterization which reads Eq. (\ref{par4}). For this parameterization, at the present time we have $c(z)=c_0$, while at the early time where $z\rightarrow\infty$, one can find $c(z)= c_0 -c_1\ln 2$.
If we consider this form for $c(z)$, we could not investigate the future behavior of $c(z)$, because it diverges when $z \rightarrow-1$. Taking derivative of Eq.\ref{par4} we have

\begin{eqnarray}\label{par4prim}
c'(z)=\frac{c_1\ln(2+z)}{(1+z)}-\frac{c_1}{(2+z)}\;.
\end{eqnarray}
Finally, inserting Eqs. \ref{par4} and \ref{par4prim} in Eq. \ref{EOS}, we obtain
\begin{eqnarray}\label{EOS4}
w_{\rm d}=\frac{2}{3}\frac{c_1(1+z)^2\left[ (1+z)+(2+z)\ln(2+z)\right]}{(2+z)\left[c_0(1+z)+c_1\ln(2+z)-c_1(1+z)\ln2 \right]}\times  \nonumber \\
\frac{1}{(1+z)^2-\left(c_0(1+z)+c_1\ln(2+z)-c_1(1+z)\ln2\right)^2}\;
\end{eqnarray}

\section{HDE models against expansion observational data}\label{sect:bg}

\begin{figure*} 
	\centering
	\includegraphics[width=5.6cm]{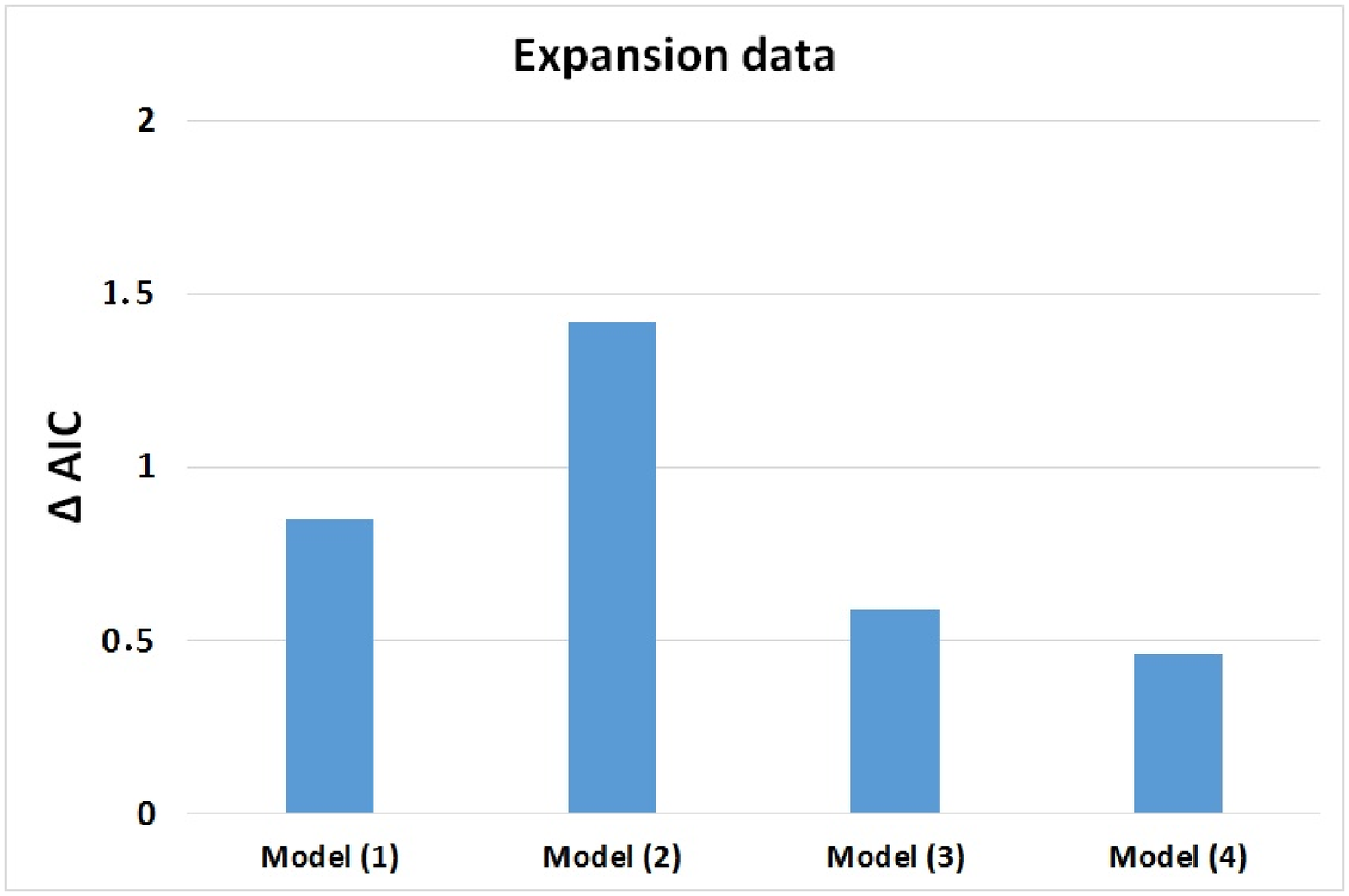}
	\includegraphics[width=5.6cm]{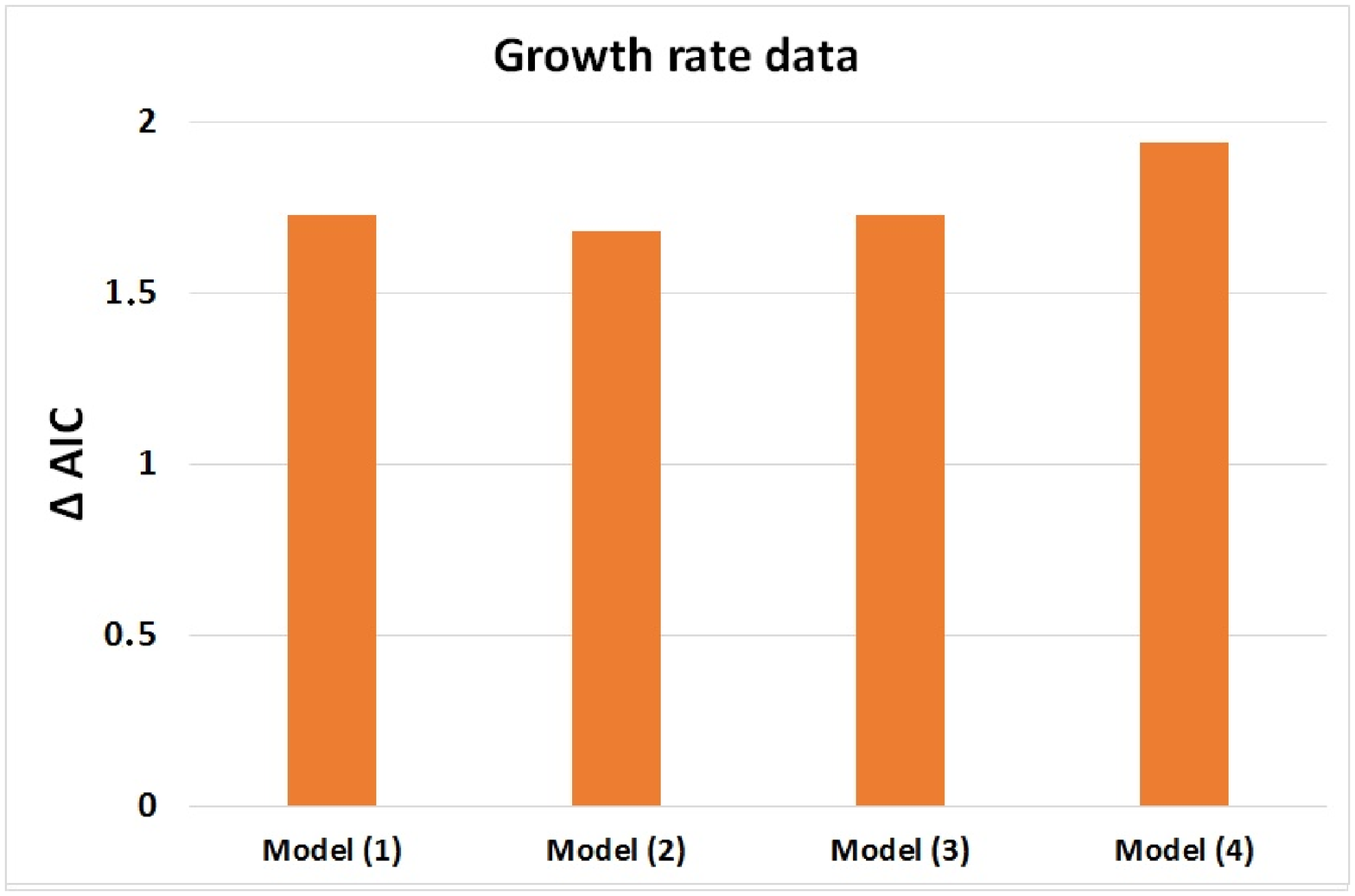}
	\includegraphics[width=5.6cm]{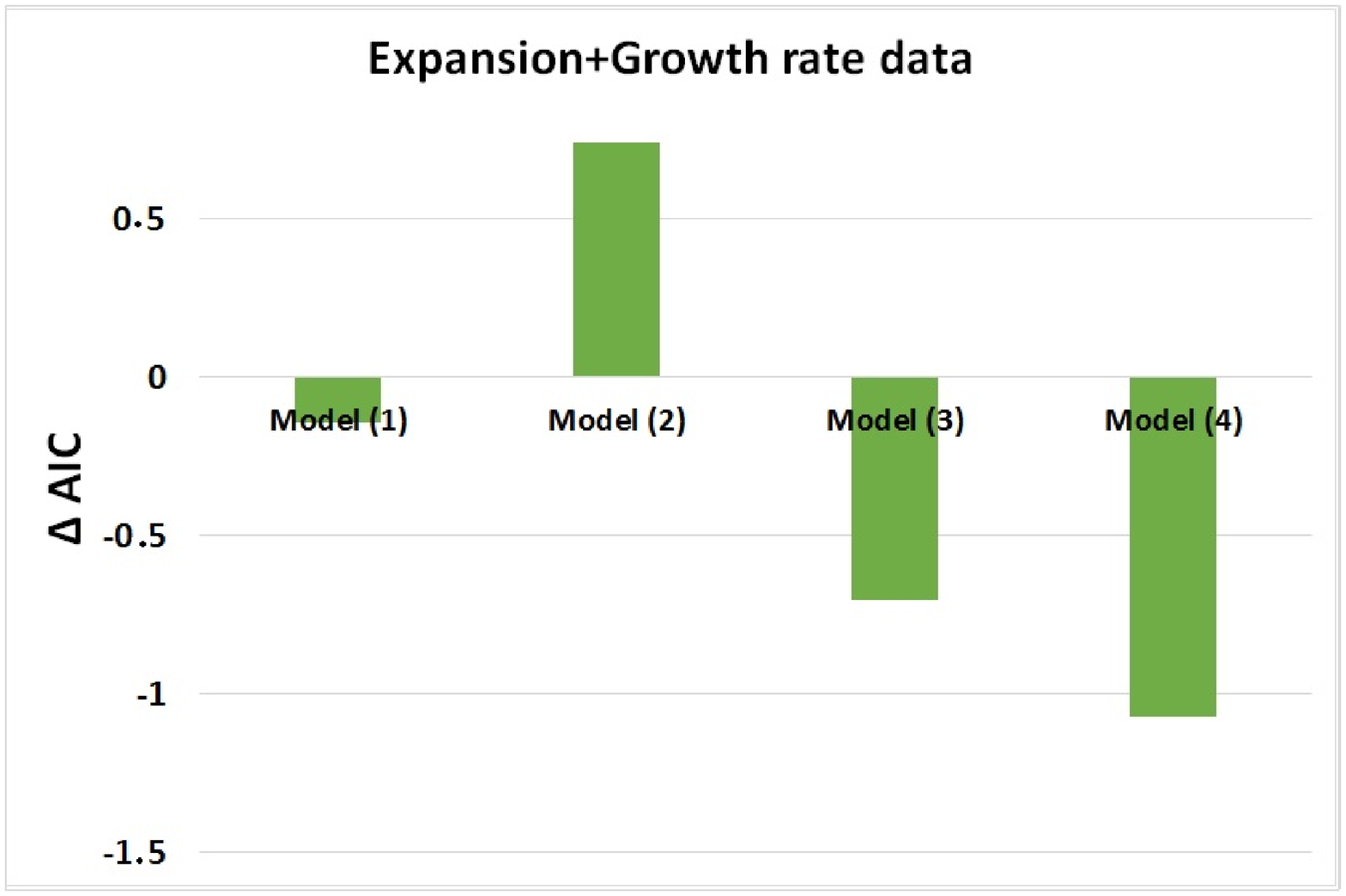}
	\includegraphics[width=5.6cm]{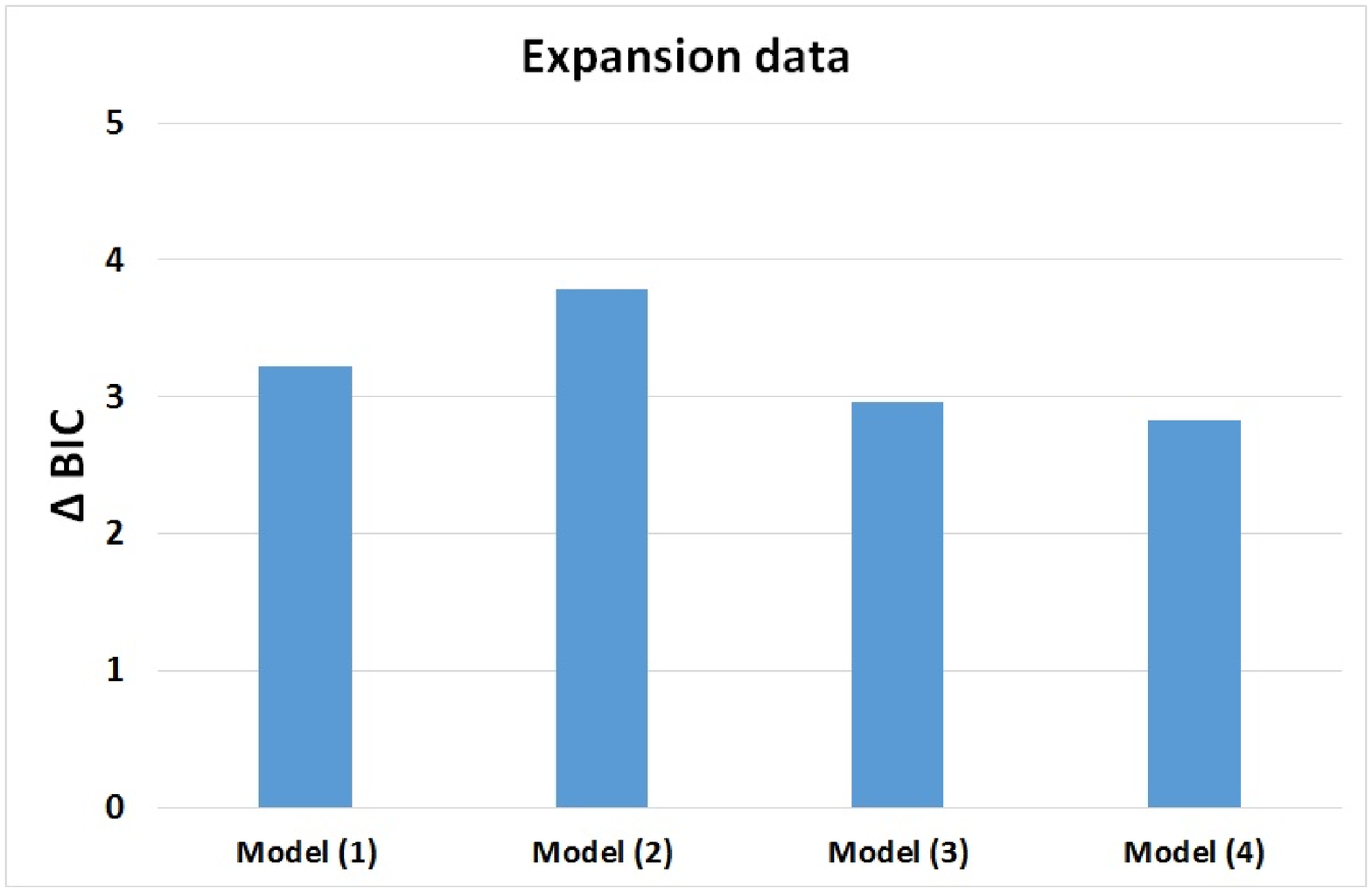}
	\includegraphics[width=5.6cm]{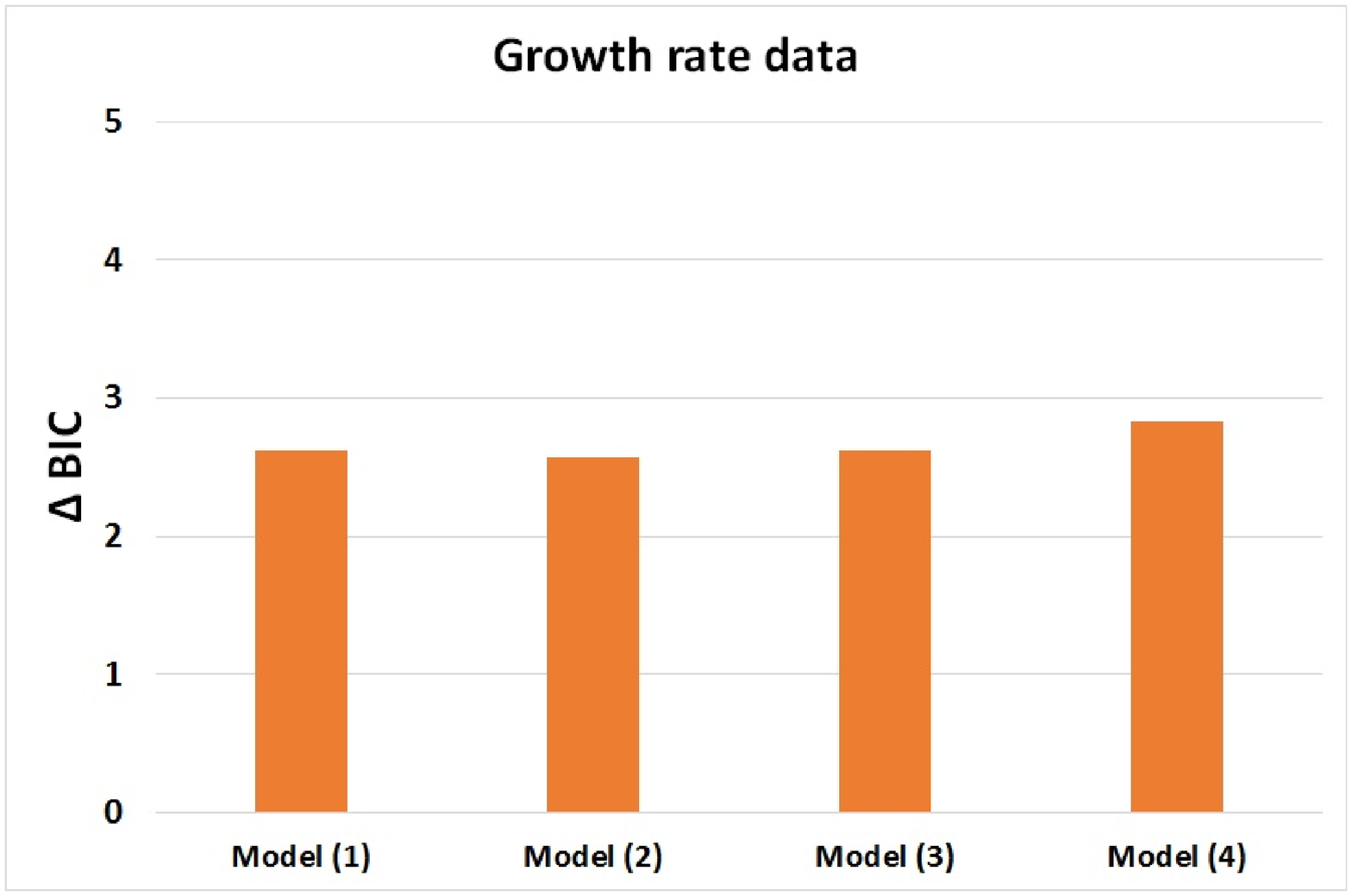}
	\includegraphics[width=5.6cm]{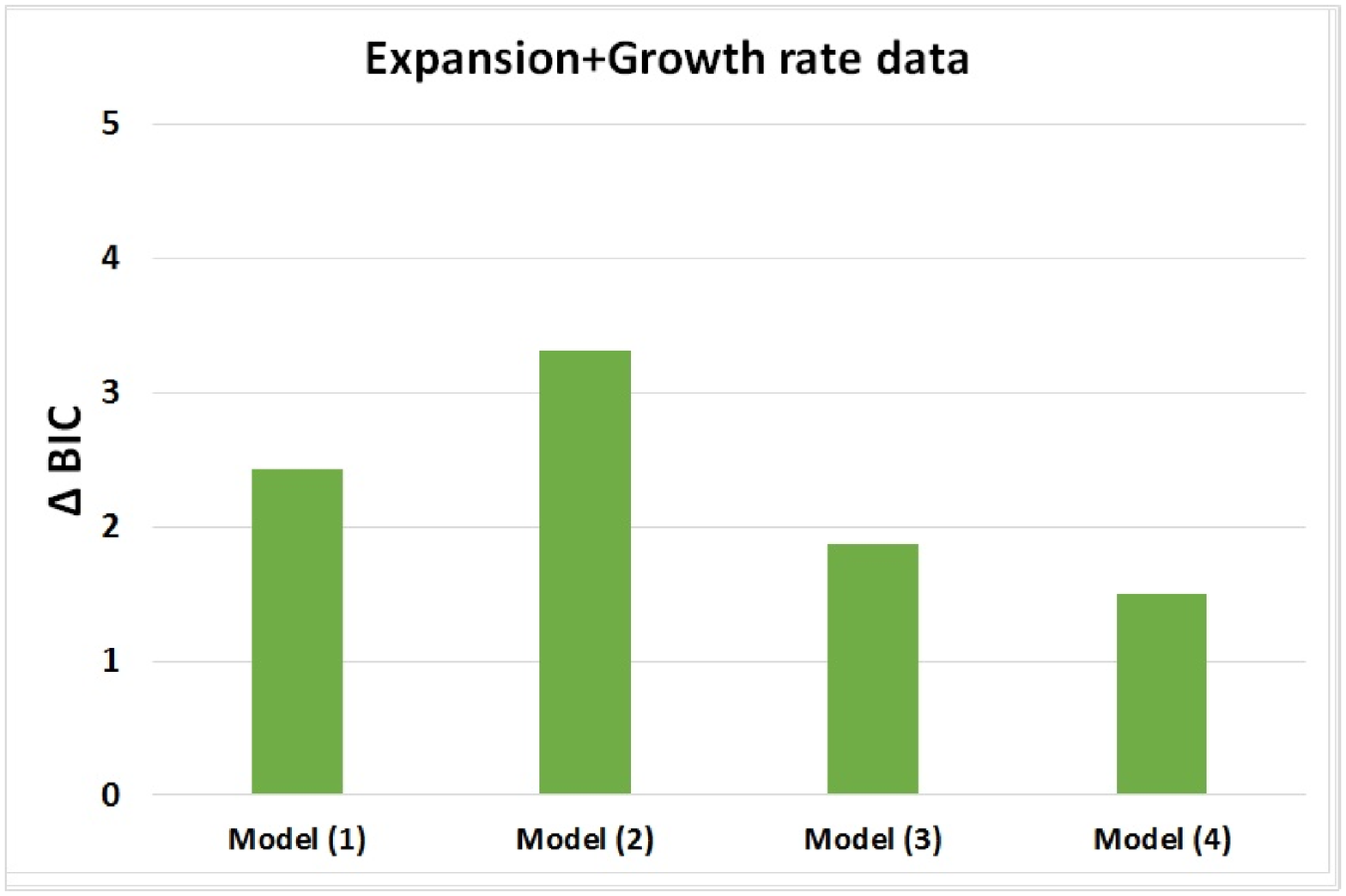}
	\caption{ The values of $\Delta {\rm AIC}$ (upper panels) and $\Delta {\rm BIC}$ (lower panels) obtained from MCMC analysis using expansion data (right panels), growth rate data (midle panels) and all of observational data (left panels) for different HDE models. }
	\label{fig:AIC}
\end{figure*}

Bellow, we investigate the HDE models using the above parameterizations for $c(z)$, against the
latest observational data in background expansion level. 
Specifically, we perform a statistical analysis using the background expansion data 
including those of SnIa (we utilize here the catalog with a binned sample data which
consists of $31$ SNe Ia events from joint light-curve analysis (JLA) \citep{Betoule:2014frx,Escamilla-Rivera:2016aca}), BAO \citep{Beutler:2011hx,Blake:2011en,Padmanabhan:2012hf,
Anderson:2012sa,Hinshaw:2012aka}, the position of the acoustic peak in the Planck CMB data \citep{Shafer:2013pxa}, BBN data point which constrains mostly $\Omega_{\rm b0}$ \citep{Serra:2009yp,Burles:2000zk}, Hubble data from the redshift evolution of cosmic chronometers \citep{Moresco:2012jh,Gaztanaga:2008xz,Blake:2012pj,Anderson:2013zyy}, and the recent data point of Hubble constant $H_0$ \citep{2018ApJ...855..136R}. Concerning the Planck CMB experiment, in our analysis we use the method of distance priors which are proposed to be a compressed likelihood to substitute the full CMB power spectrum analysis  \citep[see][]{Bond:1997wr,Efstathiou:1998xx,Wang:2007mza,Chen:2018dbv}. In these studies, CMB data are incorporated by using constraints on parameters$(R,l_a,\Omega_bh^2)$ instead of using the full CMB power spectra. It has been shown that measuring the parameters$(R,l_a,\Omega_bh^2)$ provide an efficient and intuitive summary of CMB data as far as dark energy constraints are concerned. Also in  REFF.\citep{Chen:2018dbv}, the authors compared the distance prior method with the full CMB power spectra analysis by constraining some dark energy models and showed that the results from both methods are in full agreement.
Considering these points, the total chi-square $\chi^2_{\rm tot}$ is written as:
\begin{equation}\label{eq:like-tot_chi}
 \chi^2_{\rm tot}({\bf p})=\chi^2_{\rm sn_{JLA}}+\chi^2_{\rm bao}+\chi^2_{\rm cmb}+\chi^2_{\rm h}+\chi^2_{\rm bbn}+\chi^2_{\rm H_0}\;,
\end{equation}
where the statistical vector ${\bf p}$ consists of the free parameters we have consider in the Markov chain Monte Carlo (MCMC) analysis. 
In this section the above parameters are $\lbrace\Omega_{\rm DM0},\Omega_{\rm b0}, h, c_1\rbrace$ for various HDE models and  $\lbrace\Omega_{\rm DM0},\Omega_{\rm b0}, h\rbrace$ for $\Lambda CDM$ model. By setting $z=0$ in Eq.\ref{e20}, it is easy to show that $c_0$ is a dependent parameter which equals $\sqrt{1-\Omega^0_{\rm m}-\Omega^0_{\rm r}}$.  Based on the Maximum Likelihood Principle, maximizing ${\cal L}_{\rm tot}({\bf p})$ or equivalently minimizing $\chi^2_{\rm tot}({\bf p})$ leads to the best compatibility between the models under study and the observational data points \citep{Trotta:2017wnx}. Thus we use MCMC analysis to find the best values of free parameters which minimize $\chi^2_{\rm tot}({\bf p})$. To compare HDE models considered in this work, we use the well known information criteria, namely 
AIC \citep{Akaike:1974} and BIC \citep{Schwarz:1974}.
In particular, AIC and BIC are given by 
\begin{eqnarray}
{\rm AIC} = \chi^2_{\rm min}+2k\;,\nonumber\\
{\rm BIC} = \chi^2_{\rm min}+k\ln N\;.
\end{eqnarray}
where $\chi^2_{\rm min}$ is the minimum value of $\chi^2_{\rm tot}$, $k$ is the number of free parameters and $N$ is the total number of observational data points. In this paper, at  the background level we have $N=79$ and $k=3$ for $\Lambda$CDM and $k=4$ for HDE models respectively. Reader can find more details about computing of the $\chi^2(\textbf{p})$ function, the MCMC analysis, the Akaike information criterion (AIC) and the Bayesian information criterion (BIC) in \citep{Mehrabi:2015hva} \citep[see also][]{Basilakos:2009wi,Hinshaw:2012aka,Mehrabi:2015kta,Mehrabi:2016exz,Malekjani:2016edh}. In the current step of our analysis we present the statistical results in the first column of Table\ref{tab:best}. We find that all the HDE models provide the values of $\chi^2_{\rm min}$ close to that of the usual $\Lambda$ cosmology. But since HDE models have one free parameter more than those of $\Lambda$CDM, we should compare their AIC and BIC values with that of in $\Lambda$CDM cosmology. We find $\Delta {\rm AIC}={\rm AIC}-{\rm AIC}_{\rm \Lambda}<2$ which  indicates that the HDE models considered in this study are consistent with the expansion data as equally as $\Lambda$CDM cosmology. In the case of BIC, since the $\Delta {\rm BIC}={\rm BIC}-{\rm BIC}_{\rm \Lambda}$ for all of HDE models is smaller than $6$, so there is no strong evidence against HDE models. Note that the HDE models defined based on the event horizon, Ricci and GO length scales have mild and strong tensions with background expansion data \citep[see][]{Akhlaghi:2018knk}. In the left panels of Fig.\ref{fig:AIC} we present the numerical values of $\Delta {\rm AIC}$ (up-left) and $\Delta {\rm BIC}$ (bottom-panel) for different HDE models studied in this work.
\begin{table*}
 \centering
 \caption{The values of $\chi^2_{\rm min}(AIC, BIC)$ for the different HDE models considered in this work. These results are based on the background expansion data (exp), growth rate data (gr) and combination of them (exp+gr).
The concordance $\Lambda$CDM model is shown for comparison.
}
\begin{tabular}{c  c  c c c c}
\hline \hline
 Model / Data & exp & gr (Homogeneous) & gr (Clustered)& exp+gr (Homogeneous) & exp+gr (Clustered)\\
\hline 
Model (1) & $72.78(80.78,90.26)$ & $7.71(17.71,22.16)$ & $7.71(17.71,22.16)$ & $79.80(89.80,102.67)$ & $79.77(89.77,102.64)$\\
\hline
Model (2)& $73.35(81.35,90.83)$&$7.66(17.66,22.11)$ &$7.65(17.65,22.10)$ &$80.68(90.68,103.55)$ &$80.88(90.88,103.75)$\\
\hline
Model (3) & $72.52(80.52,90.00)$&$7.71(17.71,22.16)$ & $7.74(17.74,22.19)$& $79.24(89.24,102.11)$& $79.17(89.17,102.04)$\\
\hline
Model (4) & $72.39(80.39,89.87)$&$7.92(17.92,22.37)$ & $7.80(17.80,22.25)$& $78.87(88.87,101.74)$& $79.15(89.15,102.02)$\\
\hline
$\Lambda$CDM & $73.93(79.93,87.04)$&$7.98(15.98,19.54)$ & $-$& $81.94(89.94,100.24)$& $-$\\
\hline \hline
\end{tabular}\label{tab:best}
\end{table*}

\begin{table*}
 \centering
 \caption{A summary of the best-fit parameters for the HDE models using expansion data.
}
\begin{tabular}{c  c  c c c c}
\hline \hline
 Model & Model (1) & Model (2) & Model (3)& Model (4) & $\Lambda$CDM\\
\hline 
$\Omega_{\rm m}^{(0)}$ & $0.2892^{+0.0047,+0.0088}_{-0.0047,-0.0085}$ & $0.2876^{+0.0049,+0.0089}_{-0.0049,-0.0085}$ & $0.2901^{+0.0048,+0.0090}_{-0.0048,-0.0085}$ &  $0.2904^{+0.0048,+0.0086}_{-0.0048,-0.0087}$ & $0.2893^{+0.0045,+0.0085}_{-0.0045,-0.0083}$\\
\hline
$ h $& $0.7057^{+0.0042,+0.0090}_{-0.0046,-0.0085}$& $0.7062^{+0.0052,+0.0095}_{-0.0052,-0.010}$ &$0.7059^{+0.0047,+0.0092}_{-0.0047,-0.0091}$ &$0.7051^{+0.0041,+0.0077}_{-0.0041,-0.0074}$ & $0.7020^{+0.0034,+0.0083}_{-0.0034,-0.0082}$\\
\hline
$c_1 $ & $-0.061^{+0.057,+0.10}_{-0.050,-0.11}$ &  $-0.085^{+ 0.073,+0.16}_{- 0.086,-0.15}$ & $0.073^{+0.049,+0.12}_{-0.064,-0.11}$& $0.25^{+0.10,+0.35}_{-0.16,-0.24}$ & --\\
\hline \hline
\end{tabular}\label{tab:bestfitbg}
\end{table*}

\begin{table*}
 \centering
 \caption{A summary of the best-fit parameters for the homogeneous HDE models using growth rate data.
}
\begin{tabular}{c  c  c c c c}
\hline \hline
 Model & Model (1) & Model (2) & Model (3)& Model (4) & $\Lambda$CDM\\
\hline 
$\Omega_{\rm m}^{(0)}$ & $0.280^{+0.037,+0.051}_{-0.019,-0.062}$ & $0.267^{+0.045+0.060}_{-0.025-0.073}$  &$0.267^{+0.046,+0.061}_{-0.024,-0.077}$ & $0.262^{+0.048,+0.065}_{-0.027,-0.079}$  & $0.278^{+0.0094,+0.011}_{-0.0094,-0.011}$\\
\hline
$ h $& $1.197^{+0.031,+2.7}_{-1.2,-1.2}$ & $3.19^{+1.4,+1.6}_{-0.30,-3.1}$ &  $0.60^{+0.33,+0.80}_{-0.43,-0.71}$ &  $2.6^{+2.7,+2.8}_{-2.3,-2.4}$ & $0.986^{+0.009,+0.012}_{-0.009,-0.012}$\\
\hline
$c_1 $ & $-0.75^{+ 0.56,+1.3}_{- 0.56,-1.2}$ &  $0.58^{+0.92,+1.3}_{-0.78,-1.4}$ & $0.14^{+0.20,+0.78}_{-0.56,-0.60}$ &  $-1.3^{+1.2,+1.5}_{-2.2,-2.7}$ & -- \\
\hline
 $ \sigma_{8}(z=0) $ & $0.693^{+0.040,+0.14}_{-0.068,-0.11}$ &  $0.814^{+0.051,+0.15}_{-0.089,-0.12}$ &$0.779^{+0.028,+0.19}_{-0.096,-0.13}$ & $0.813^{+0.028,+0.15}_{-0.076,-0.10}$ & $0.803^{+0.021,+0.034}_{-0.021,-0.034}$\\
\hline \hline
\end{tabular}\label{tab:bestfitgr1}
\end{table*}

\begin{table*}
 \centering
 \caption{A summary of the best-fit parameters for the clustered HDE models using growth rate data.
}
\begin{tabular}{c  c  c c c }
\hline \hline
 Model & Model (1) & Model (2) & Model (3)& Model (4) \\
\hline 
$\Omega_{\rm m}^{(0)}$ & $0.268^{+0.049,+0.062}_{-0.020,-0.085}$& $0.270^{+0.044,+0.060}_{-0.023-0.075}$ & $0.277^{+0.038,+0.053}_{-0.021,-0.064}$ & $0.276^{+0.040,+0.054}_{-0.021,-0.067}$  \\
\hline
$ h $&$0.89^{+0.60,+0.76}_{-0.80,-0.85}$ & $0.97^{+0.65,+1.3}_{-0.92,-0.95}$ &  $1.02^{+0.63,+1.3}_{-0.73,-1.2}$ & $1.78^{+0.86,+1.5}_{-0.86,-1.7}$  \\
\hline
$c_1 $ & $0.04^{+1.0,+1.4}_{-0.82,-0.86}$&$0.14^{+0.40,+1.5}_{-0.80,-0.96}$  & $2.9^{+2.3,+2.4}_{-3.0,-3.1}$ & $0.26^{+0.40,+0.70}_{-0.40,-0.71}$  \\
\hline
 $ \sigma_{8}(z=0) $ &$0.781^{+0.022,+0.23}_{-0.11,-0.13}$ &  $0.780^{+0.034,+0.14}_{-0.073,-0.11}$ & $0.6606^{+0.0067,+0.12}_{-0.056,-0.067}$& $0.755^{+0.023,+0.071}_{-0.040,-0.063}$  \\
\hline \hline
\end{tabular}\label{tab:bestfitgr2}
\end{table*}

\begin{table*}
 \centering
 \caption{A summary of the best-fit parameters for the homogeneous HDE models using expansion + growth rate data.
}
\begin{tabular}{c  c  c c c c}
\hline \hline
 Model & Model (1) & Model (2) & Model (3)& Model (4) & $\Lambda$CDM\\
\hline 
$\Omega_{\rm m}^{(0)}$ & $0.2823^{+0.0082,+0.016}_{-0.0082,-0.016}$ & $0.2822^{+0.0084,+0.016}_{-0.0084,-0.016}$  & $0.2822^{+0.0079,+0.015}_{-0.0079,-0.015}$ &$0.2831^{+0.0079,+0.015}_{-0.0079,-0.016}$ &$0.2861^{+0.0078,+0.014}_{-0.0078,-0.014}$ \\
\hline
$ h $ &$0.7135^{+0.0092,+0.019}_{-0.0092,-0.017}$ &  $0.7125^{+0.0074,+0.020}_{-0.011,-0.017}$  & $0.7146^{+0.0092,+0.018}_{-0.0092,-0.017}$ &$0.7132^{+0.0094,+0.018}_{-0.0094,-0.017}$ & $0.7049^{+0.0067,+0.012}_{-0.0067,-0.012}$\\
\hline
$c_1 $ &$-0.086^{+0.066,+0.12}_{-0.053,-0.12}$ & $-0.123^{+0.12,+0.19}_{-0.060,-0.27}$ & $0.098^{+0.066,+0.13}_{-0.066,-0.13}$ &  $0.33^{+0.29,+0.38}_{-0.42,-0.48}$& -- \\
\hline
 $ \sigma_{8}(z=0) $ & $0.742^{+0.020,+0.041}_{-0.020,-0.037}$ & $0.745^{+0.019,+0.041}_{-0.021,-0.037}$ &$0.741^{+0.020,+0.038}_{-0.020,-0.038}$  & $0.741
^{+0.022,+0.045}_{-0.022,-0.044}$ & $0.749^{+0.022,+0.043}_{-0.022,-0.043}$\\
\hline \hline
\end{tabular}\label{tab:bestfittot1}
\end{table*}

\begin{table*}
 \centering
 \caption{A summary of the best-fit parameters for the clustered HDE models using expansion + growth rate data.}
\begin{tabular}{c  c  c c c }
\hline \hline
 Model & Model (1) & Model (2) & Model (3)& Model (4) \\
\hline 
$\Omega_{\rm m}^{(0)}$ &$0.2828^{+0.0081,+0.016}_{-0.0081,-0.016}$ & $ 0.2844^{+0.0084,+0.016}_{-0.0084,-0.016}$ & $0.2823^{+0.0077,+0.015}_{-0.0077,-0.015}$& $0.2854^{+0.0079,+0.015}_{-0.0079,-0.015}$  \\
\hline
$ h $ &$0.7130^{+0.0079,+0.017}_{-0.0091,-0.016}$ & $0.7083^{+0.0092,+0.018}_{-0.0092,-0.017}$& $0.7144^{+0.0083,+0.016}_{-0.0083,-0.016}$& $0.7068^{+0.0080,+0.015}_{-0.0080,-0.016}$  \\
\hline
$c_1 $ &$-0.084^{+ 0.056,+0.11}_{- 0.056,-0.12}$ & $-0.055^{+0.075,+0.15}_{-0.075,-0.16}$ &$0.095^{+0.052,+0.10}_{-0.052,-0.10}$ &  $0.08^{+0.21,+0.34}_{-0.25,-0.31}$ \\
\hline
 $ \sigma_{8}(z=0) $ & $0.746^{+0.022,+0.042}_{-0.022,-0.044}$ & $0.748^{+0.021,+0.042}_{-0.021,-0.041}$ & $0.745^{+0.021,+0.040}_{-0.021,-0.042}$  & $0.749^{+0.020,+0.038}_{-0.020,-0.040}$ \\
\hline \hline
\end{tabular}\label{tab:bestfittot2}
\end{table*}

\begin{figure} 
	\centering
	\includegraphics[width=8cm]{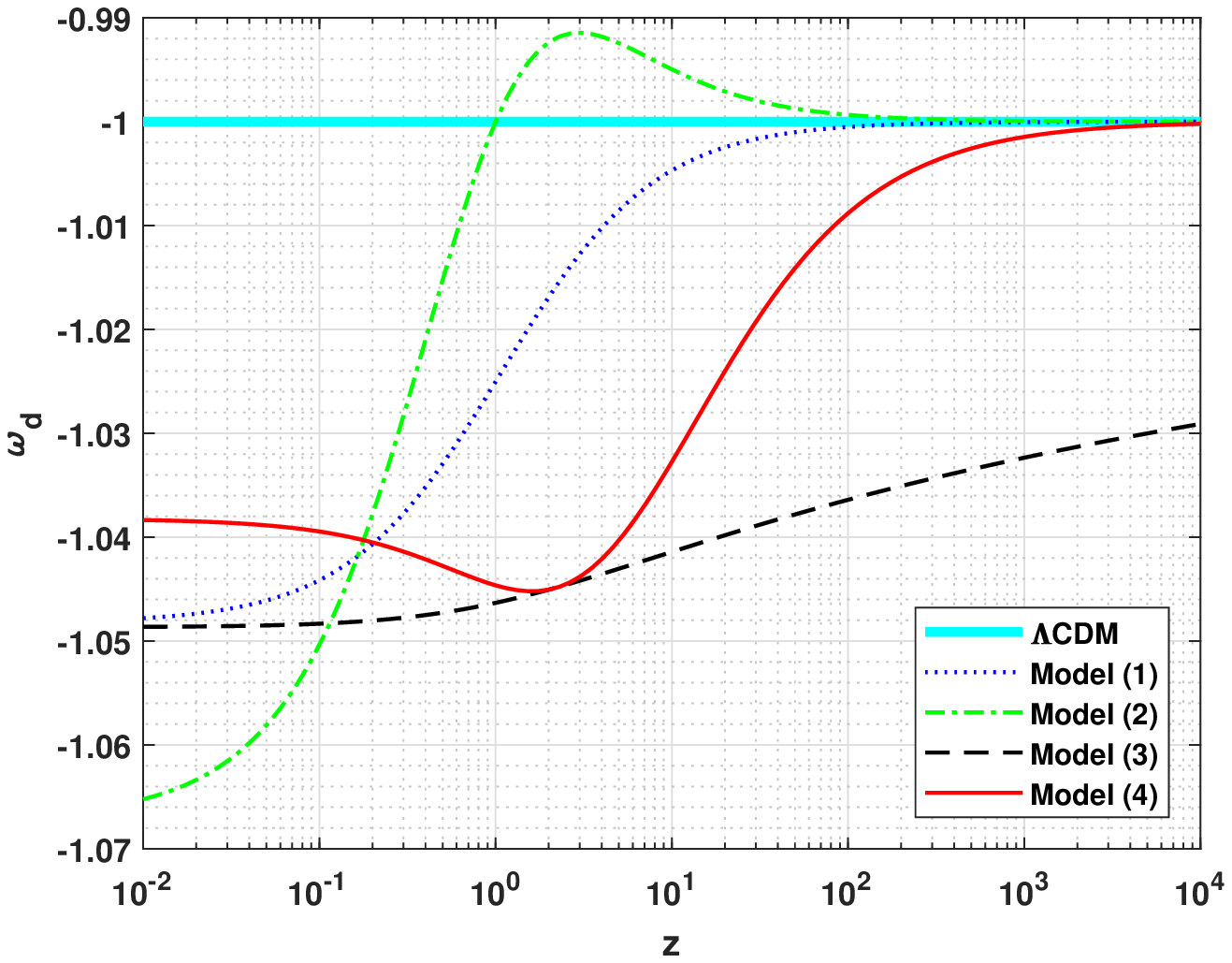}
	\includegraphics[width=8cm]{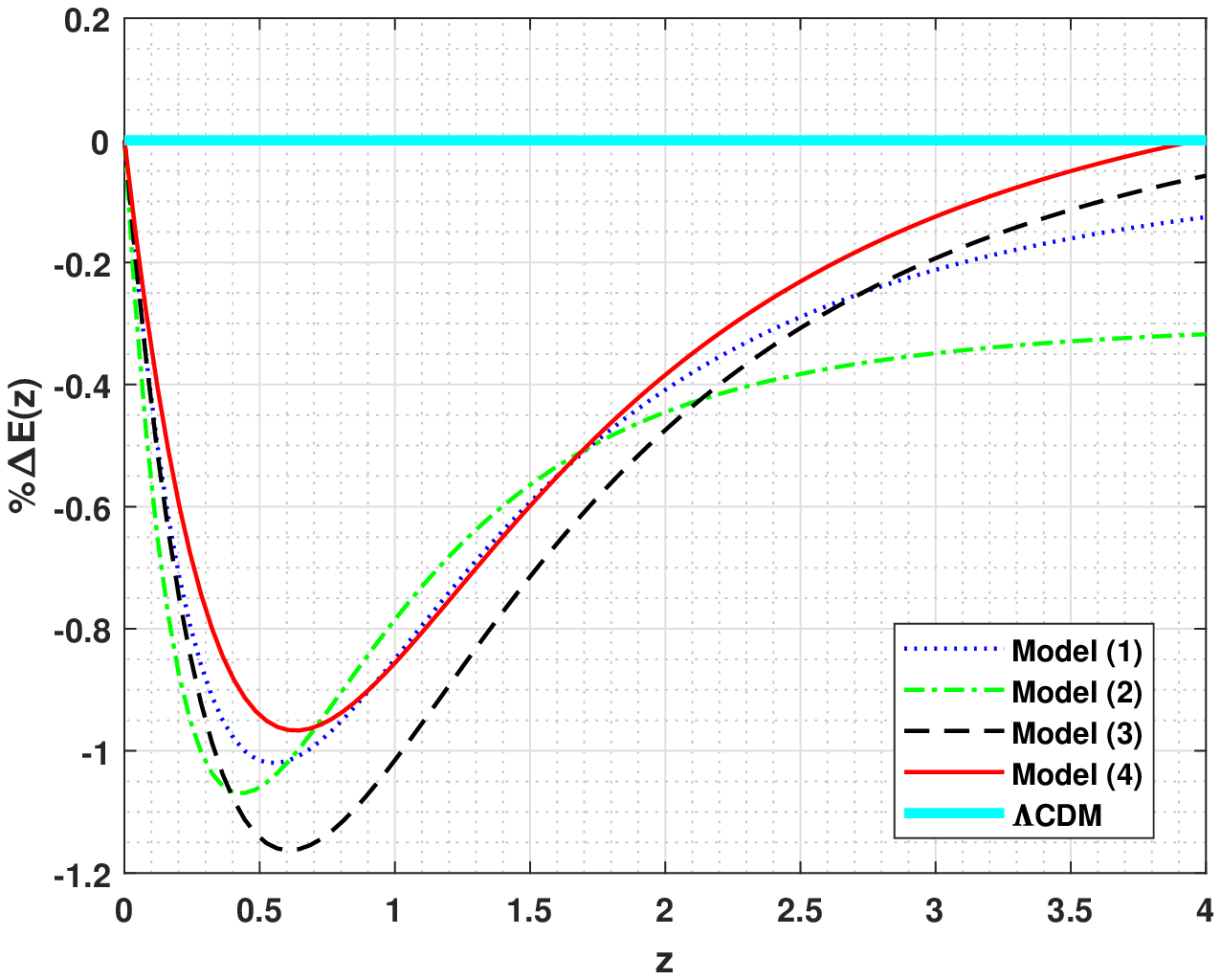}
	\includegraphics[width=8cm]{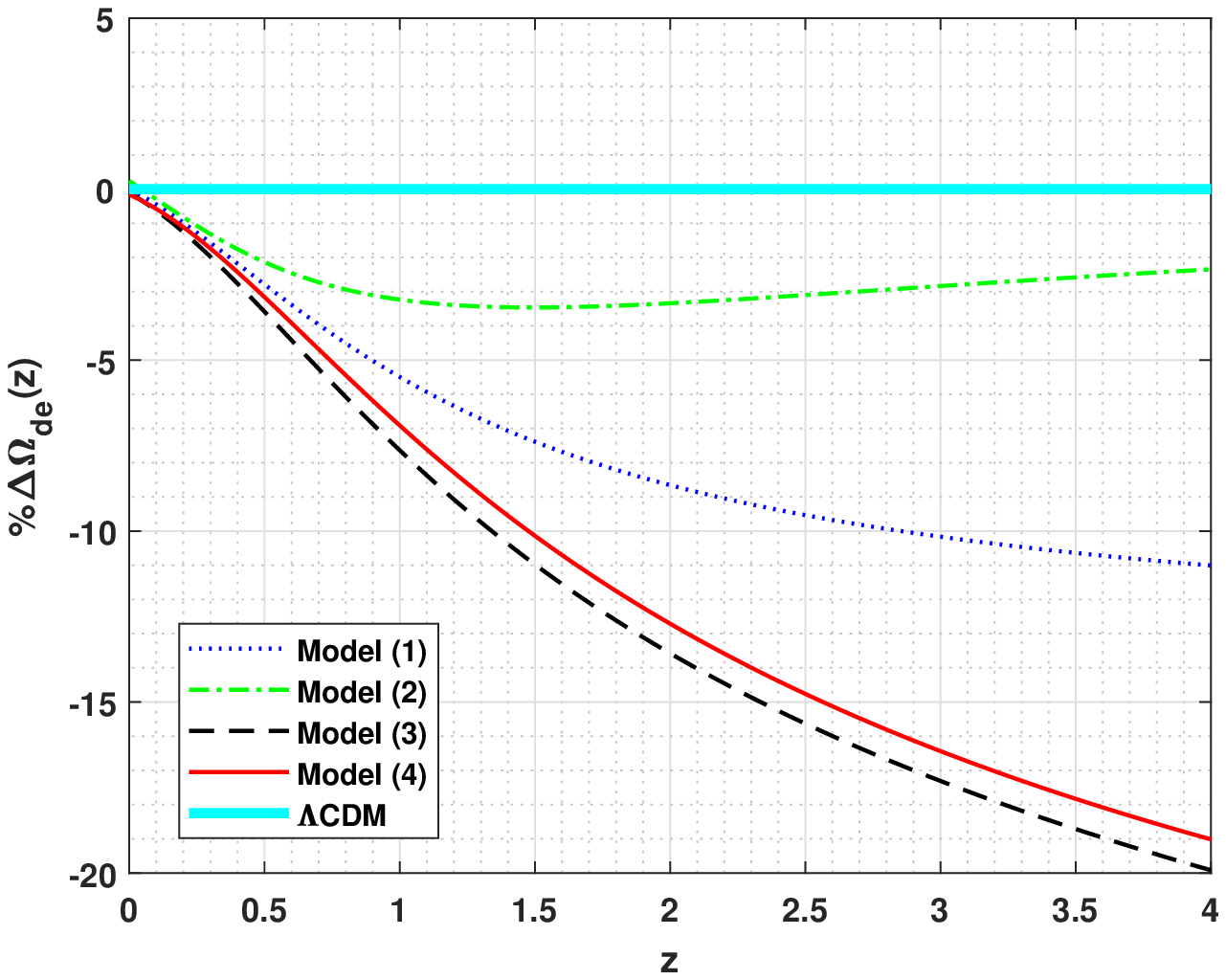}
	
	\caption{ The redshift evolution of different cosmological quantities, namely dark energy EoS parameter $w_{\rm d}(z)$ ( top panel), $\Delta E(\%)$ (middle panel) and $\Delta \Omega_{\rm d}(\%)$ ( bottom panel).
		The different HDE models are characterized by the colors and line-types presented in the inner panels of the figure.}
	\label{fig:back}
\end{figure}

\begin{figure} 
	\centering
	\includegraphics[width=8cm]{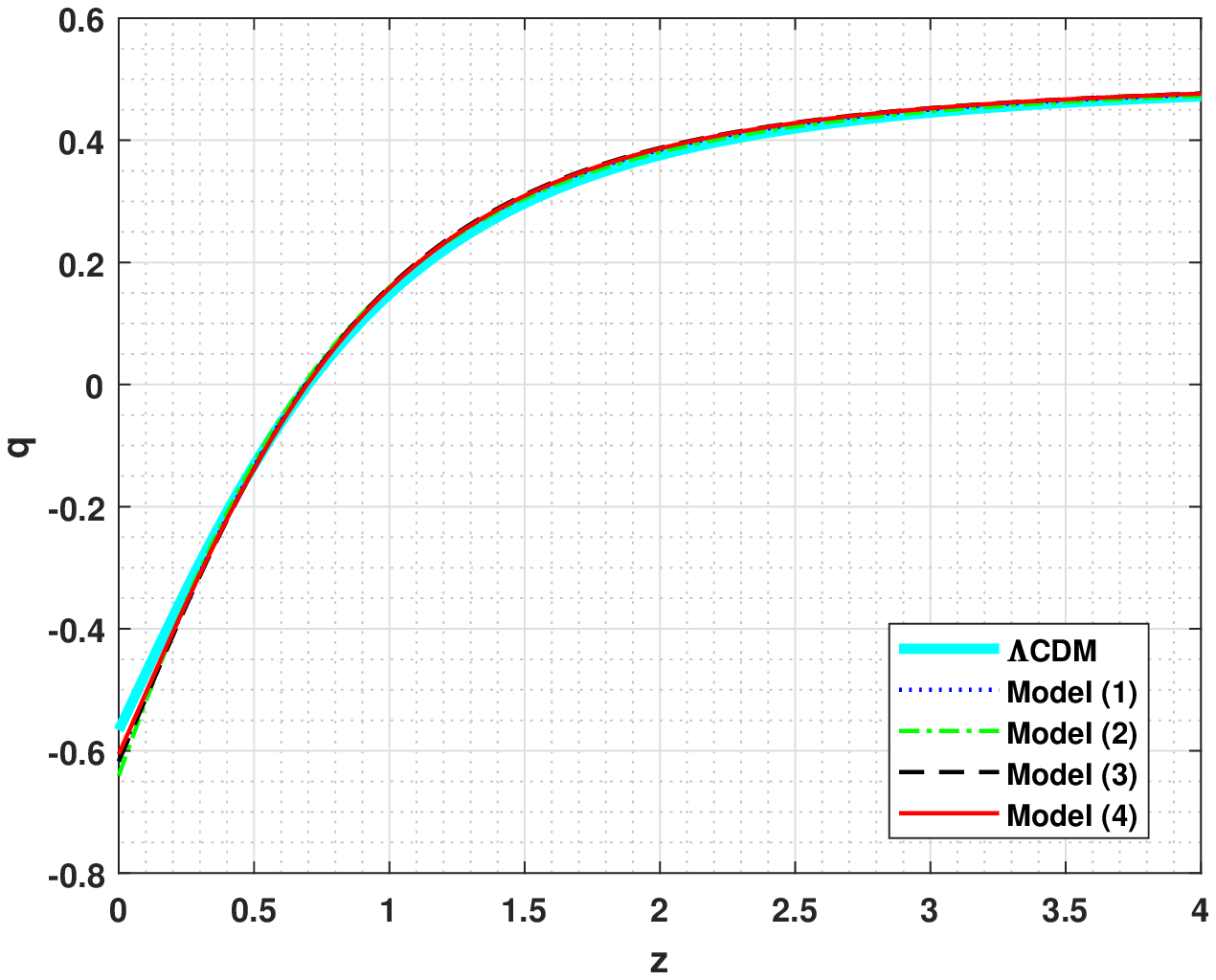}
	\caption{ The redshift evolution of deceleration parameter $q(z)$ using best fit parameters obtained from MCMC analysis. The different HDE models are characterized by the colors and line-types presented in the inner panel of the figure.}
	\label{fig:cz}
\end{figure}

We also present the best fit values and mean $1\sigma$ and $2\sigma$ errors of free parameters for different HDE models in Table (\ref{tab:bestfitbg}). Using these best fit parameters in Fig.\ref{fig:back} we plot the redshift evolution of main cosmological quantities in background level including the EoS parameter, $w_{\rm d}$ (upper panel), the percentage of relative difference of Hubble parameter $\Delta E(\%)=[(E-E_{\rm \Lambda})/E_{\rm \Lambda}]\times 100$ (middle panel) and the percentage of relative difference of DE density parameter $\Delta \Omega_{\rm d}(\%)=[(\Omega_{\rm d}-\Omega_{\rm \Lambda})/\Omega_{\rm \Lambda}]\times 100$ (lower panel) for different HDE models considered in this work. We observe that the present value of EoS of HDE models are smaller than $-1$. This result shows that the EoS parameter of HDE models evolves in the phantom region ($w_{\rm d}<-1$) at low redshifts. Particularly, in the case of Model (2), the EoS parameter of DE evolves from quintessence regime at higher redshifts and enter the phantom regime at $z\sim1$. While for other HDE models considered in this work, the EoS parameter varies slowly in the phantom regime until relatively higher redshifts. Notice that at very high rdshifts, the EoS parameter of Models (1, 2 \& 4) coincide to $w_{\Lambda}=-1$. From the middle panel of Fig.\ref{fig:back}, we observe that the value of $E_{\rm d}(z)$ for HDE models is smaller than $E_{\rm \Lambda}(z)$ at low redshifts. This means that compared to $\Lambda$ cosmology, in the presence of HDE models the universe experiences smaller expansion rate at lower redshifts. The maximum value of relative difference $\Delta E$ for different HDE models varies between $ 0.9-1.2\%$ and occurs at redshifts $z \sim 0.5$. Lastly, from the bottom panel of Fig.\ref{fig:back} we find that the value of $\Omega_{\rm d}$ for all HDE models is smaller than $\Omega_{\rm \Lambda}$. 
Now let us switch to the deceleration parameter $q$, which is one of the main cosmological parameters in background level representing the phase expansion of the universe. It is defined in terms of scale factor, as

 \begin{eqnarray}
q=-\dfrac{a\ddot{a}}{\dot{a}^2}=-1-\dfrac{\dot{H}}{H^2}\;.\label{qqq}
 \end{eqnarray}

Following standard lines, one can obtain \citep{Rezaei:2017yyj}
\begin{eqnarray}
\dfrac{\dot{H}}{H^2}=-\dfrac{3}{2}(1+w_{\rm d}\Omega_{\rm d})\;.\label{qq}
\end{eqnarray}
Replacing Eqs.(\ref{omeghde} \& \ref{EOS}) in Eq.\ref{qq} and inserting the result in Eq.\ref{qqq} we would have the deceleration parameter $q$ in HDE models as follows

\begin{eqnarray}
q=\dfrac{1}{2}-\dfrac{3}{2}\dfrac{c^2(z)}{E^2(z)}-\dfrac{c'(z)c(z)}{E^2(z)}\;.\label{q}
\end{eqnarray}
Now using best-fit values of free parameters we plot the evolution of deceleration parameter as a function of $z$ in Fig.\ref{fig:cz}. We observe that in all cases, including that of $\Lambda$CDM, universe starts to accelerated expansion ($q<0$) at $z_{\rm tr}\sim0.7$. Similar results for transition redshift from early decelerated to current accelerated expansion in the framework of DE and modified gravity theories \citep{Capozziello:2014zda,Capozziello:2015rda,Farooq:2016zwm,Rezaei:2017yyj}. As expected, at high redshifts, the epoch at which DE has no significant effect on the evolution of universe, $q$ tends to $1/2$.

\section{HDE models against Growth rate  observational data }\label{growth}
 In this section, we investigate the linear growth of matter perturbations in the presence of HDE cosmologies. In order to find the effects of DE on the linear growth of matter fluctuations we have two different scenarios which have been widely investigated in literature \citep{ArmendarizPicon:1999rj,Garriga:1999vw,ArmendarizPicon:2000dh,Abramo2007,Abramo:2008ip,Basilakos:2009mz,Mehrabi:2015hva,Mehrabi:2015kta,Malekjani:2016edh,Rezaei:2017yyj}. We limit our analysis to sub-horizon scales, where the results of Pseudo Newtonian dynamics
are well consistent with those of General Relativity paradigm \citep[see ][]{Abramo2007}. In the first scenario, clustering DE,  perturbations of DE can grow same as matter perturbations \citep[see also][]{Abramo:2008ip,Batista:2013oca,Batista:2014uoa}. For this approach, DE perturbations are affected by the negative pressure which implies that the amplitude of DE perturbations is smaller than dark matter fluctuations. In the other scenario, the homogeneous DE case, DE perturbations cannot grow significantly in sub-Hubble scales. In homogeneous DE scenario we have $\delta_{\rm d}\equiv 0$ and only the corresponding non-relativistic matter is allowed to cluster. For both of these approaches, we refer the reader to follow our previous articles \citep{Mehrabi:2015kta,Malekjani:2016edh,Rezaei:2017yyj} in which we have provided the basic differential equations which describe the evolution of matter and DE perturbations. Concerning the initial conditions, we use those provided by \citep{Batista:2013oca} \citep[see also][]{Malekjani:2016edh,Rezaei:2017hon}. In fact using these initial conditions we verify that matter perturbations always stay in the linear regime.
Now we can obtain the evolution of fluctuations ($\delta_{\rm m}, \delta_{\rm d}$) and using them we can calculate the growth rate $f(z)={d\ln{\delta_{\rm m}}}/{d\ln{a}}$ of large scale structures in the presence of different HDE models considered in this work. We also calculate the rms mass variance at $R=8h^{-1}Mpc$ as $\sigma_8(z)=\frac{\delta_{\rm m}(z)}{\delta_{\rm m}(z=0)}\sigma_8(z=0)$ in HDE models to compare the theoretical prediction of $f(z)\sigma_8(z)$ quantity in HDE cosmologies with growth rate observational data measured from redshift space distortion (RSD) of galaxies.
In this way we perform another likelihood analysis using the growth rate ($f(z)\sigma_8$) data. In this step we have $N=18$ (number of data points) and $k=5$ for HDE models (${\bf p}=\lbrace\Omega_{\rm DM0},\Omega_{\rm b0}, h, c_1, \sigma_8\rbrace$). The numerical results form $\chi^2_{\rm min}$ analysis obtained in this step are shown in the second and third columns of Table (\ref{tab:best}) respectively for homogeneous and clustered HDE models. Furthermore, we present the relevant AIC and BIC values in the parentheses. We observe that for all HDE models $\Delta {\rm AIC}<2$ (see also middle-up panel of Fig.\ref{fig:AIC}) which implies, as well as $\Lambda$CDM, all the HDE models are consistent with the growth rate data. On the other hand, for all models we have $\Delta {\rm BIC}<6$ (see also middle-bottom panel of Fig.\ref{fig:AIC}) which supports the results obtained from AIC criteria. Thus using growth rate data only, we find that there is no strong evidence against HDE models compare to standard $\Lambda$CDM cosmology.
Same as previous section we present the best fit values and mean $1\sigma$ and $2\sigma$ errors of free parameters for different HDE models in Tables (\ref{tab:bestfitgr1} \& \ref{tab:bestfitgr2}) respectively for homogeneous and clustered DE scenarios. From the best fit value of free parameters and their relatively large errors obtained in this step, we find that growth rate data cannot solely put strong constraints on DE parameters, especially on parameter $h$ which has the greatest error value.

We now perform an overall likelihood analysis using the expansion data combined with growth rate ones. In this case, the total chi-square $\chi^2_{\rm tot}$ is given by

\begin{equation}\label{eq:like-tot_chi}
 \chi^2_{\rm tot}({\bf p})=\chi^2_{\rm sn_{JLA}}+\chi^2_{\rm bao}+\chi^2_{\rm cmb}+\chi^2_{\rm h}+\chi^2_{\rm bbn}+\chi^2_{\rm H_0}+\chi^2_{\rm gr}\;,
\end{equation}

\begin{figure} 
	\centering
	\includegraphics[width=9cm]{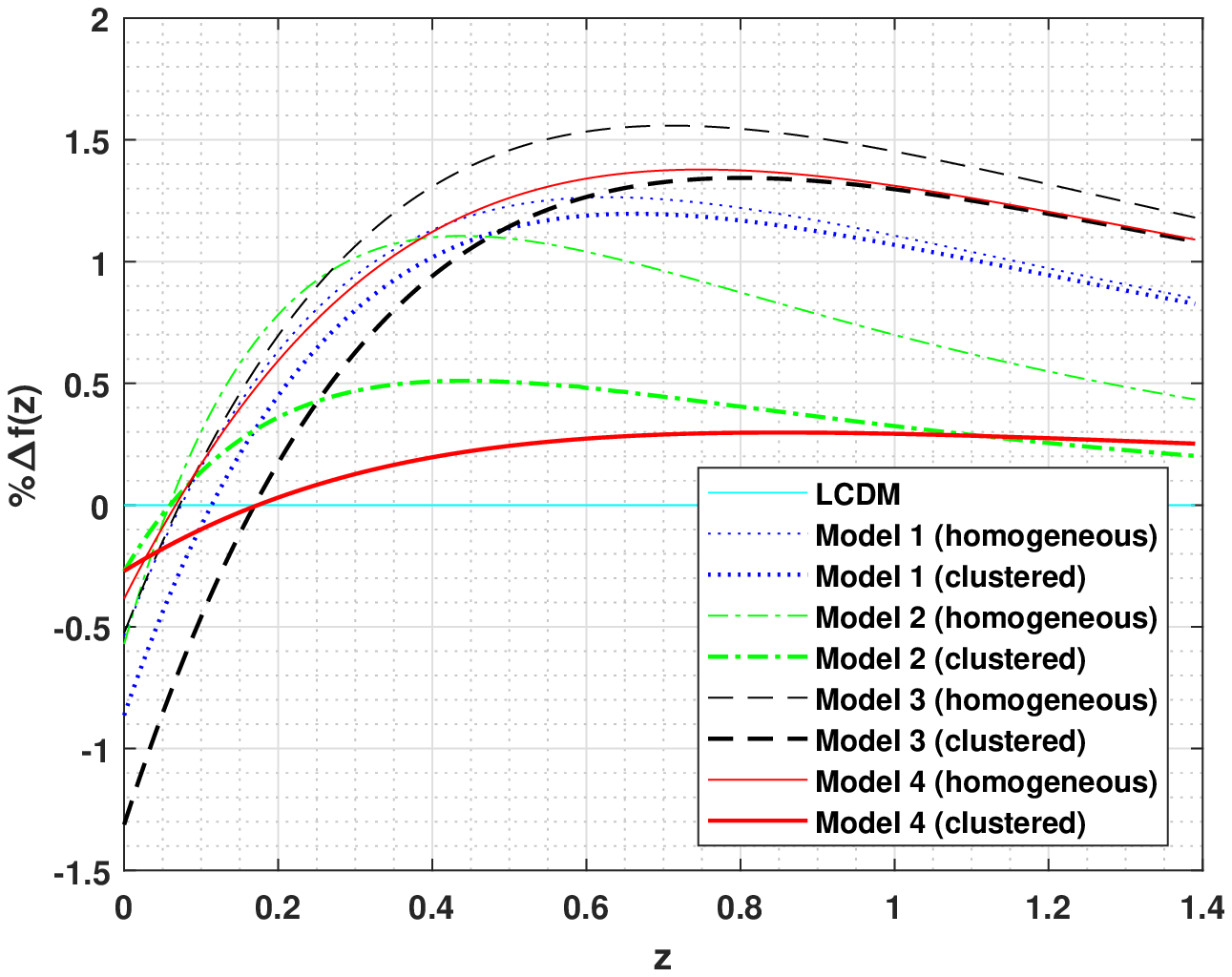}
	\includegraphics[width=9cm]{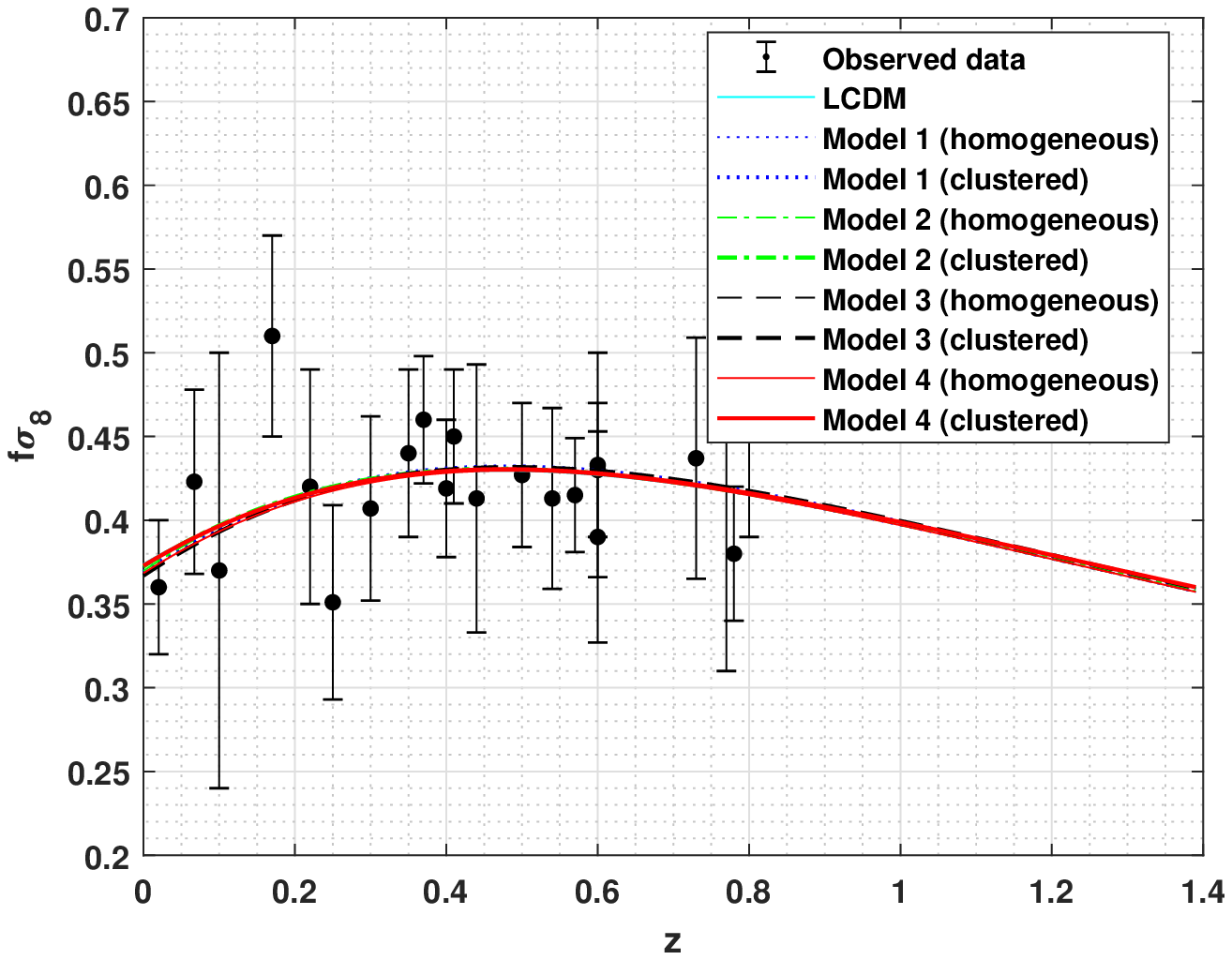}
	\caption{ The corresponding fractional difference $\Delta f(\%)=100\times [f(z)-f_{\rm \Lambda}(z)]/f_{\rm \Lambda}(z)$ (upper panel) and comparison of the observed and theoretical evolution of the growth rate $f (z)\sigma_8 (z)$ as a function of redshift $z$.  
The different HDE models are characterized by the colors and line-types presented
in the inner panels of the figure}.  
	\label{fig:growthfunction}
\end{figure}

As mentioned before, the vector ${\bf p}$ contains the free parameters of the particular cosmological model. In this step, the relevant parameters are $\lbrace\Omega_{\rm DM0},\Omega_{\rm b0}, h, c_1,\sigma_8\rbrace$, so we have $k=5$ and $N=97$. Same as the previous step, we consider two different homogeneous and clustered HDE scenarios. The results of our analysis for different HDE models, are shown in Table \ref{tab:best} (two last columns) and Tables (\ref{tab:bestfittot1} \& \ref{tab:bestfittot2}) for homogeneous and clustered DE cases respectively. 
We observe that performing overall likelihood using both background and growth rate data leads to relatively smaller values for $\Delta {\rm AIC}$ and $\Delta {\rm BIC}$. This result is more significant for Models (1,3 \& 4)where we have $\Delta {\rm AIC} < 0$ . This result indicates that the value of AIC parameter in these HDE models is smaller than the same parameter in concordance $\Lambda$CDM cosmology. However, the difference is smaller than $2$ and in this comparison we can not reject the $\Lambda$CDM model. Comparing the results obtained from combined analysis with those of Sect. \ref{sect:hde} we conclude that in the both of homogeneous and clustered DE scenarios, adding background data to the growth rate ones, leads to smaller $1\sigma$ and $2\sigma$ errors. Also, the best fit values of free parameters for different HDE models are coming closer to those found for $\Lambda$CDM model. Using the best fit values of cosmological parameters presented in Tables (\ref{tab:bestfittot1} \& \ref{tab:bestfittot2}), we plot the fractional difference growth rate $f(z)$ with respect to that of $\Lambda$CMD model ($\Delta f(\%)=100\times [f(z)-f_{\rm \Lambda}(z)]/f_{\rm \Lambda}(z)$) in the upper panel of Fig.\ref{fig:growthfunction}. In the lower panel, the observed $f(z)\sigma_8(z)$ is compared to the theoretical predicted growth rate for different HDE models. We see that all HDE models are fitted to observational growth rate data as well as concordance $\Lambda$CDM universe.

We observe that for all HDE models, the evolution of $\Delta f$ has an maximum at low redshifts. As expected, this feature in the evolution of $\Delta f$ is related to the evolution of $\Delta E$. Indeed, we verify that higher values of the normalized Hubble parameter $E(z)$ correspond to smaller values of the growth rate. Thus, when $\Delta E$ has a minimum we expect the growth rate $\Delta f$ to have a maximum and vice versa (see middle panel of Fig. \ref{fig:back} and upper panel of Fig.\ref{fig:growthfunction}). At high redshifts the relative difference $\Delta f\rightarrow 0$, since at high redshifts, the role of DE becomes negligible and universe is matter dominated, namely $\delta_{\rm m}\propto a$. Thus the growth rate for all HDE models as well as $\Lambda$CDM model tends 
to unity and therefore the relative difference tends to zero.  
The present value of $\Delta f$, for homogeneous ( clustered) Model (1) is 
$\sim [-0.5\%,1.2 \%]$ ( $[-0.8\%,1.2 \%]$). In the case of Model (2) we have 
$\Delta f\sim [-0.5\%,1.1 \%]$ and $[-0.2\%,0.5 \%]$ for homogeneous and clustered 
DE respectively. For homogeneous (clustered) Model (3) the relative deviation lies in the interval $\sim [-0.5\%,1.6\%]$ ( $[-1.3\%,1.4 \%]$). Finally, for homogeneous ( clustered) Model (4) we obtain $\Delta f\sim [-0.4\%, 1.4 \%]$ ( $[-0.3\%, 0.3 \%]$).

\section{Conclusions}
\label{conlusion}
In this work we investigated the cosmological properties of holographic DE with varying $c^2$ term in the context of Hubble distance, $H_0$ as the IR cutoff. We considered four different well known parameterizations to describe the evolution of $c(z)$ parameter. After putting constraints on the free parameters of the models using background observational data, we studied the behavior of the basic cosmological quantities include $w_d(z), \Omega_d(z), E(z)$ and $q(z)$ in the presence of different HDE models. In the perturbation level we used latest growth rate data and consider homogeneous and clustered DE scenarios. We showed that all of HDE models under study, are well fitted to cosmological data like $\Lambda$CDM model, both at background and perturbation levels.
In particular, our main results may be summarized as follows:

\textit{(i)} Initially, using the latest background observational data 
we performed a likelihood analysis for different HDE cosmologies in the context of MCMC method. Based on this analysis we placed constraints on the free parameters of models and we showed that all HDE models are consistent with the  background data as equally as concordance $\Lambda$CDM universe. Using the best fit values we plotted the evolution of $w_{\rm d}, \Delta E$ and $\Delta \Omega_{\rm d}$ in Fig.\ref{fig:back}. We found that the present value of $w_{\rm d}$ in all HDE models is in the phantom region. At  $z\sim1$, Model (2) crosses the phantom line $w=-1$ while rest of the HDE models remains in the phantom regime until relatively higher redshifts. At early enough times, the EoS parameter of HDE models (1,2 and 4) mimics the constant EoS $w_{\Lambda}=-1$ of $\Lambda$CDM cosmology. We found that the Hubble parameter in HDE cosmologies is $ \sim 0.9-1.2\%$ smaller than the $\Lambda$CDM model at low redshifts. We also showed that in HDE, cosmologies the universe changes its phase from decelerating to accelerating expansion at $z_{\rm tr}\sim0.7$ which in $1\sigma$ error is consistent with observations \citep[see also][]{Capozziello:2014zda,Capozziello:2015rda,Farooq:2016zwm,Rezaei:2017yyj}.

\textit{(ii)} 
 We performed a statistical analysis using the growth rate data in order to put constraints on free parameters of models. In this step we obtained  best fit parameters with relatively large error bars. This means that the growth rate data could not put tight constraints on the cosmological parameters. However the results for $\chi^2_{min}$, $\Delta AIC$ and $\Delta BIC$ showed that HDE models considered in this work are well consistent with recent growth rate data.  

\textit{(iii)} 
Finally, we performed a joint statistical analysis using the combined expansion and growth rate data in order to compare the models and put constraints on their free parameters. Based on the best fit parameters obtained in this step, we plotted the evolution of the fractional difference $\Delta f(\%)$ for HDE models. The maximum value of $\Delta f(\%)$ occurs at relatively low redshifts, when the role of DE becomes more significant, while the differences among HDE models are negligible at higher redshifts.We found that the absolute value of the difference between AIC (BIC) criteria of HDE models with that of obtained in $\Lambda$CDM cosmology is smaller than $2$ ($4$). Hence we concluded that the HDE models with time varying model parameter defined on Hubble length are well fitted to observational data as equally as concordance $\Lambda$CDM model. In an other word, by  current cosmological data there is no even weak evidence against HDE models proposed in this work. 

\section{Acknowledgements}
The work of MR has been supported financially by Research Institute for Astronomy \& Astrophysics of Maragha (RIAAM) under research project No. 1/5440-38.


 \bibliographystyle{apsrev4-1}
  \bibliography{ref}

\end{document}